\documentclass[aps,floatfix,prl,letterpaper]{revtex4}
\usepackage{graphicx}
\usepackage{dcolumn}
\usepackage{bm}
\usepackage{subfigure}
\usepackage{epstopdf}
\usepackage{natbib}
\usepackage{amsmath} 
\usepackage{color} 
\usepackage{longtable} 
\usepackage{tabularx}
\usepackage[abs]{overpic}

\begin{document}
\title{Elastohydrodynamics of contact in adherent sheets}
\author{Andreas Carlson$^1$, Shreyas Mandre$^2$ and L. Mahadevan$^1$} 
\affiliation{$^1$Paulson School of Engineering and Applied Sciences, Department of Physics, \\Wyss Institute for Biologically Inspired Engineering,  \\Harvard University, 29 Oxford Street, Cambridge, MA 02138}
\affiliation{$^2$School of Engineering, Brown University, Providence, USA.}

\date{\today}
\begin{abstract}{Adhesive contact between a thin elastic sheet and a substrate in a liquid environment arises in a range of biological, physical and technological applications. By considering the dynamics of this process that naturally couples fluid flow, long wavelength elastic deformations and microscopic adhesion, and solving the resulting partial differential equation numerically, we uncover the short-time dynamics of the onset of adhesion and the long-time dynamics of a steady propagating adhesion front.  Simple scaling laws corroborate our results for characteristic waiting-time for adhesive contact, as well as the speed of the adhesion front. A similarity analysis of the governing partial differential equation further allows us to determine the shape of a fluid filled bump ahead of the adhesion zone. Finally, our analysis yields the boundary conditions for the apparent elastohydrodynamic contact line, generalizing the well known conditions for static elastic contact while highlighting how microscale physics regularizes the dynamics of contact.}
\end{abstract}
 \maketitle
\section{Introduction}
Adhesion of thin elastic films to a substrate occurs in a range of problems in physical chemistry \cite{Bell:1978}, in biology \cite{Giannone:2007}, and in {engineering} \cite{Chaudhury2005,Tong,Bengtsson1996,Turner:2004,weihua:2000,Rieutord2005,Radisson:2013,Navarro:2013}. Experimental and theoretical studies of adhesion focus on the microscale physical chemistry of bond formation \cite{Bell:1978,Bell:1984}, and the mesoscale mechanics of elastic deformation of the film, and squeeze flow in the gap \cite{Cantat:1998,Chaudhury2013EPL, Mani:2012, Hosoi:2004, Lister2013PRL,Leong,Pihler:2012,michaut, Bruyn:2015,fitton:2004}. In the neighborhood of the contact line itself, all these effects are potentially important - short range adhesive interactions affect the squeeze flow, which can lead to a high pressure that deforms the film and affects the dynamics of adhesion in return. However, theoretical models of soft adhesion usually focus only on the elastic and adhesive aspects of the process, typically neglecting the hydrodynamics \cite{Rieutord:2014, Springman:2008,Springman:2009}. An important exception is \cite{Rieutord2005}  where energetic arguments were used to determine the steady bonding speed of silicon wafers coming into contact and the self-similar shape of the contact zone, but without a complete consideration of the microscopic nature of adhesion or the transient dynamics of contact.

\begin{figure}
\centering
\includegraphics[width=0.80\linewidth]{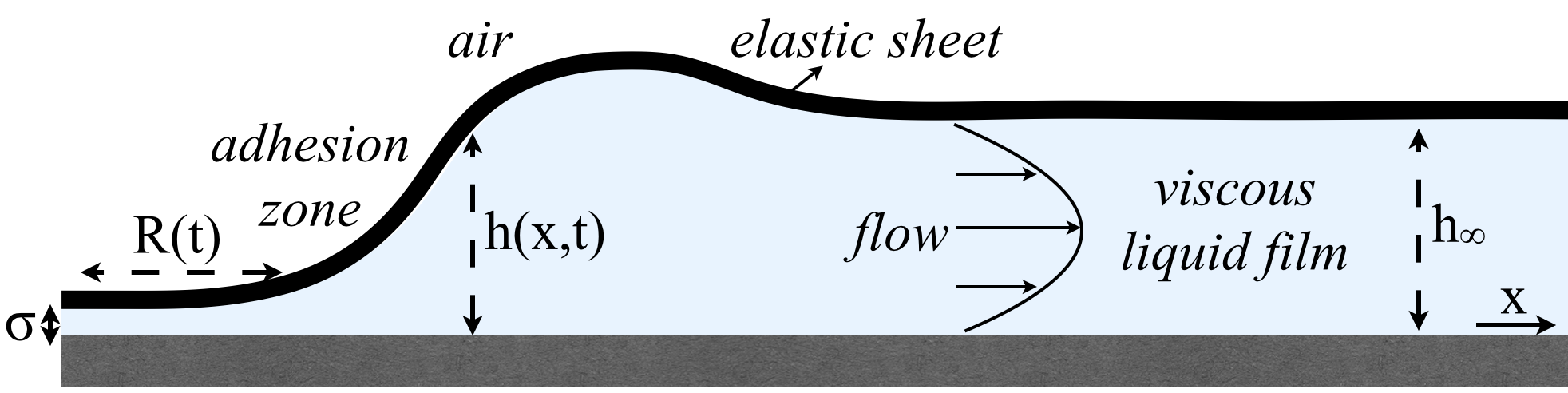}
\caption{Sketch showing variables. An elastic sheet with a thickness $b$, which is separated from a solid substrate by a thin viscous film of height $h=h(x,t)$. The sheet is adhered to the substrate at the left side at a height $h=\sigma$ and has at the right side a far-field height $h_{\infty}$. The adhesion pressure $\Phi_h$ brings the sheet towards the substrate leading to the formation of an adhesion zone $R(t)$ that propagates towards the right. \label{Maysketch}}
\end{figure}

Here we consider the complete transient elastohydrodynamics of soft adherent sheets with the goal of characterizing the onset of adhesion, the generic properties of the moving contact zone once it reaches a steady state. We also determine the appropriate boundary conditions at the effective contact line (Fig. 1),  thus generalizing the boundary conditions for a static elastic contact line, first studied theoretically and experimentally nearly a century ago \cite{Obreimoff1930}. These questions are analogous to those raised a long time ago for liquid-vapor-solid contact lines, which have an accompanying rich literature \cite{degennes85,Bonn2009, Snoeijer2013}. Just as in liquid-vapor-solid contact lines, elastohydrodynamic contact is associated with a near-singular mathematical behavior. 

\section{Mathematical model}
\begin{figure}
\centering
\includegraphics[width=1.0\linewidth]{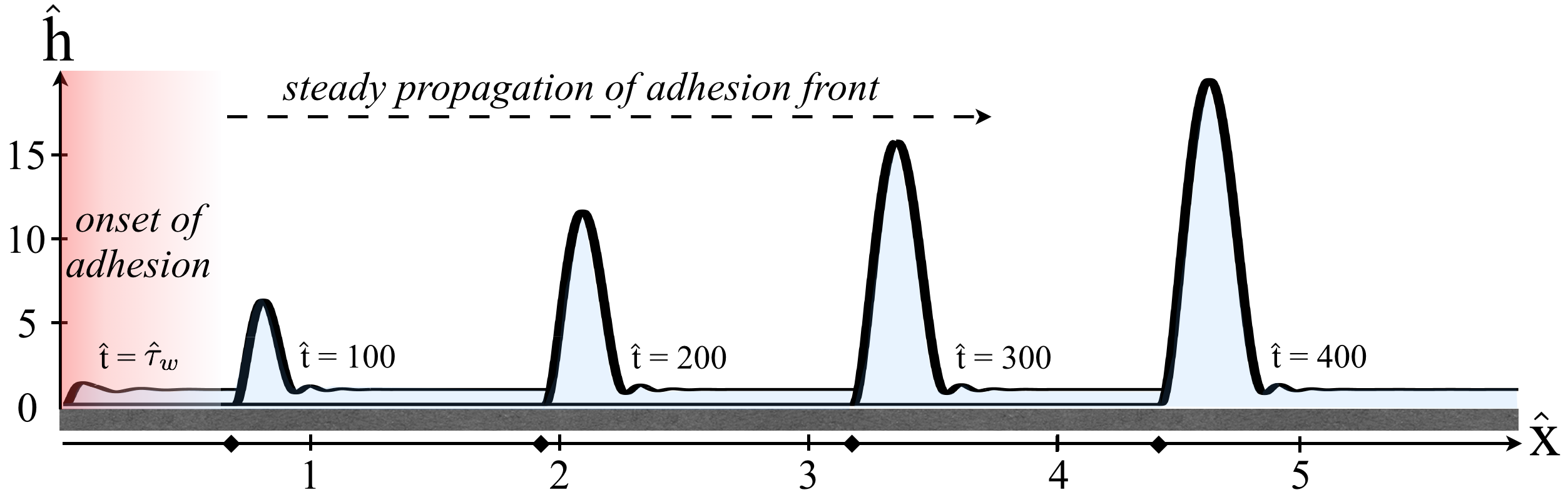}
\caption{Elastohydrodynamics of adhesion.  A thin and long elastic sheet that is initially separated from a solid substrate by a viscous film is brought in contact at the left edge by a microscopic adhesion potential. The shape of the sheet follows from the evolution equation (\ref{tfAdh}), starting with an initial condition given as $\hat h(\hat x,\hat t=0)=1+0.1\times(1-\tanh(25\hat x)^2)$ that trigger adhesion. The results are presented in non-dimensional form for five different points in time $\hat t=[\hat\tau_w=47, 100, 200, 300, 400]$, with $\hat B=5\times 10^{-7}$, $\epsilon=10^{-2}$ and $\hat \sigma=0.12$. After a short transient phase associated with the formation of a contact zone given by the waiting-time $\hat \tau_w=\tau_w/\tau_{\mu}=\epsilon^{-2} (3\hat B)^{\frac{1}{2}}$, the adhesion zone propagates in time displacing the fluid into a bump. The position of the apparent elastic contact line $\hat R(\hat t)\equiv\max(\hat h_{\hat x\hat x}(\hat t,\hat x))$ is indicated by the diamond shaped markers along the $\hat x$-axis, equidistantly spaced in time. Note that the x-axis is compressed by a factor $\approx 1/\epsilon=100$.\label{sketch}}
\end{figure}
\subsection{Elasticity}
We begin with a description of the elasticity of a long thin and very wide sheet of length $L$ [m], thickness $b [m] \ll L$, separated from a solid substrate by a viscous thin film of height $h=h(x,t)$ [m] and $h\ll L$, where we denote the dimensions of the variables within [$\cdot$]. The elastic sheet is attracted to a substrate by an intermolecular adhesion potential (Fig. 1), consisting of a van der Waals attraction and a strong short range-repulsion $\Phi(h)=\frac{A}{3}\left(\frac{1}{2 h^2}-\frac{\sigma^6}{8 h^8}\right)$ \cite{Israelachvili2011}, with $A$ [J] the Hamaker constant. Here $\sigma$ [m] is the equilibrium height where the attraction and the repulsion pressure balance; evaluating the energy at this minimum yields the effective solid-solid surface tension $\gamma = \Phi(\sigma)=A/8\sigma^2$ [J/m$^2$]. Then the shape of the sheet is determined by the balance of transverse forces on the sheet and yields the following equation for long wavelength elastic deformations  \cite{Landau1986}
\begin{equation}
p=Bh_{xxxx}- \Phi_h.
\label{eq:p}
\end{equation}
Here $p$ is the local pressure along the sheet, $B$ [J] is its bending stiffness and $(\cdot )_x\equiv \partial (\cdot )/\partial x$, and we have further assumed that $h_x^2\ll1$, so that we can neglect any geometrical nonlinearities in the shape of the sheet. 

For a static sheet, the pressure $p=0$. In this case, multiplying (\ref{eq:p}) by $h_x$ and integrating by parts, or alternatively using the principle of virtual work, yields the boundary conditions at the static elastic contact line   $h(R)=h_x(R)=0, h_{xx}(R)=(\frac{2\gamma}{B})^{\frac{1}{2}}$ derived and experimentally verified nearly a century ago by Obreimoff \cite{Obreimoff1930}, see Supplemental Information (SI).

\subsection{Hydrodynamics} When the sheet is allowed to move in response to the attractive adhesive interaction which brings it into proximity with the substrate, the interstitial fluid begins to be squeezed out.   On small scales, the inertia of the fluid and the sheet is dominated by viscous forces in the gap. In low Reynolds number limit, i.e. $Re=\rho h_{\infty}^2 c/(L\mu) \ll 1$ with $\rho$ [kg/m$^3$] the liquid density, $h_{\infty}$ the initial film height, $c$ [m/s] the adhesion speed along the substrate, and $\mu$ [Pa$\cdot$ s] the fluid viscosity, the fluid flow field, ${\bf u} = (u(x,y,t), v(x,y,t))$ is governed by the incompressible two-dimensional Stokes flow equations $\nabla p = \mu \Delta {\bf u}, \nabla \cdot {\bf u}=0$.   Evaluating the different terms in the Stokes equations show that to leading order in $\epsilon \equiv  h_{\infty}/L\ll 1$ the thin film flow can be described by the lubrication equations \cite{Batchelor}
\begin{equation}
 -p_x+\mu u_{yy}=0,~~p_y=0,~~
 u_x+v_y=0.\label{continuity}
\end{equation}
Using the no-slip boundary conditions on the two solid surfaces
 \begin{eqnarray}
u(x,0)=0,~~~u(x,h)=0,\nonumber \\
v(x,0)=0,~~~v(x,h)=h_t,
\end{eqnarray}
to integrate (\ref{continuity}) yields $u(x,y)=\frac{p_x}{2\mu}\left(y^{2}-yh\right)$. On substituting this result into the continuity equation together with (\ref{eq:p}) yields a sixth order nonlinear partial differential equation for the thin film height \cite{Oron:1997}
\begin{equation}
h_t-\left(\frac{h^3}{12\mu}\left(Bh_{xxxx}- \Phi_h\right)
_x \right)_x=0.\label{tfAdh}
\end{equation} 

\subsection{Scaling, non-dimensional numbers and boundary conditions}Scaling (\ref{tfAdh}) using the following definitions for the dimensionless variables;
\begin{equation}
\hat h\equiv \frac{h}{h_{\infty}}, \hat x\equiv \frac{x}{L}, \hat p\equiv \frac{ph_{\infty}^3}{A}, \hat t\equiv t/\tau_{\mu}\equiv \frac{tA}{12\mu h_{\infty}^3}
\label{eq:scalvar}
\end{equation}
give three non-dimensional numbers: $\hat B\equiv B h_{\infty}^4/AL^4$, the ratio between elastic bending pressure and adhesion pressure; $\hat \sigma\equiv(\sigma/h_{\infty})$, the ratio between the equilibrium adhesion height and the initial film height; and the aspect ratio $\epsilon\equiv h_{\infty}/L$. All our results in the text are given in both dimensional and scaled forms, but our figures show only the non-dimensional results scaled according to (\ref{eq:scalvar}).

In the context of experiments on wafer bonding \cite{Rieutord2005,Navarro:2013} with $\mu=2\times 10^{-5}$ Pa$\cdot$s, $h_{\infty}=100$ nm, $L=10 cm$, $A= 10^{-19}$ N$\cdot$m \cite{Israelachvili2011}, $\sigma\in [12-24]$ nm,  $B=10^{-4}-10$ N$\cdot$m, we find that $\epsilon=10^{-6}\ll1$, $\hat \sigma \in [0.06,0.12]$, $\hat B \in [10^{-9} - 10^{-4})$. Furthermore, since the bonding speed $c\in 0.1-2$ cm/s \cite{Bengtsson1996,Rieutord2005,Navarro:2013}, the Reynolds number $Re=\rho h_{\infty}^2 c/(\mu L)\approx10^{-7}\ll1$ justifying the lubrication approximation (\ref{tfAdh}). We note that since $\hat B \ll 1$ and multiplies the highest order derivative in (\ref{tfAdh}) we expect a boundary layer to form near the apparent contact line. Indeed, the size of this boundary layer $l_c$ can be estimated by balancing the bending pressure $Bh_{xxxx} \sim B/l_c^4$ and the attractive van der Waals pressure $A/3h^3$, as $h\approx \sigma$ so that $l_c\approx \sigma (3B/A)^{\frac{1}{4}}$ and is the length of the inner adhesion zone (see SI). Furthermore, we note that the slope near the apparent contact line scales as $h_x\approx \sigma/l_c\approx (A/3B)^{\frac{1}{4}}\ll1$, justifying our long wavelength approximation. 

In order to complete the problem formulation, we need to prescribe six boundary conditions at the two edges of the sheet. Assuming that the sheet is free of forces, torques and exposed to an ambient fluid environment at constant pressure, this implies that at either end, 
\begin{equation}
h_{xx}|_{x=0,L}=h_{xxx}|_{x=0,L}=p|_{x=0,L}=0.\label{eq:bcs}
\end{equation}
Our mathematical model provides a general and compact transient formulation of the entire elastohydrodynamic adhesion phenomenon that includes the microscopic physics, extending and complementing previous steady-state analyses \cite{Bengtsson1996,weihua:2000,Rieutord2005}. 

\section{Analysis}
To understand the behavior of the solution of the initial boundary value problem we solve equation (\ref{tfAdh}-\ref{eq:bcs}) numerically by using a second-order finite difference method with a spatial discretization $\delta x/L=1/400-1/1600$ and use a Gear method \cite{Gear} for the adaptive-time marching.  In Fig. \ref{sketch}, we show the results obtained by solving (\ref{tfAdh}) numerically, delineating three regimes, starting with the onset of adhesion at short times, an intermediate time regime as the fluid filled bump is formed and a long time regime associated with the steady motion of the adhesive front $R(t)$ and the fluid bump. 
\begin{figure}
\centering
\includegraphics[width=0.70\linewidth]{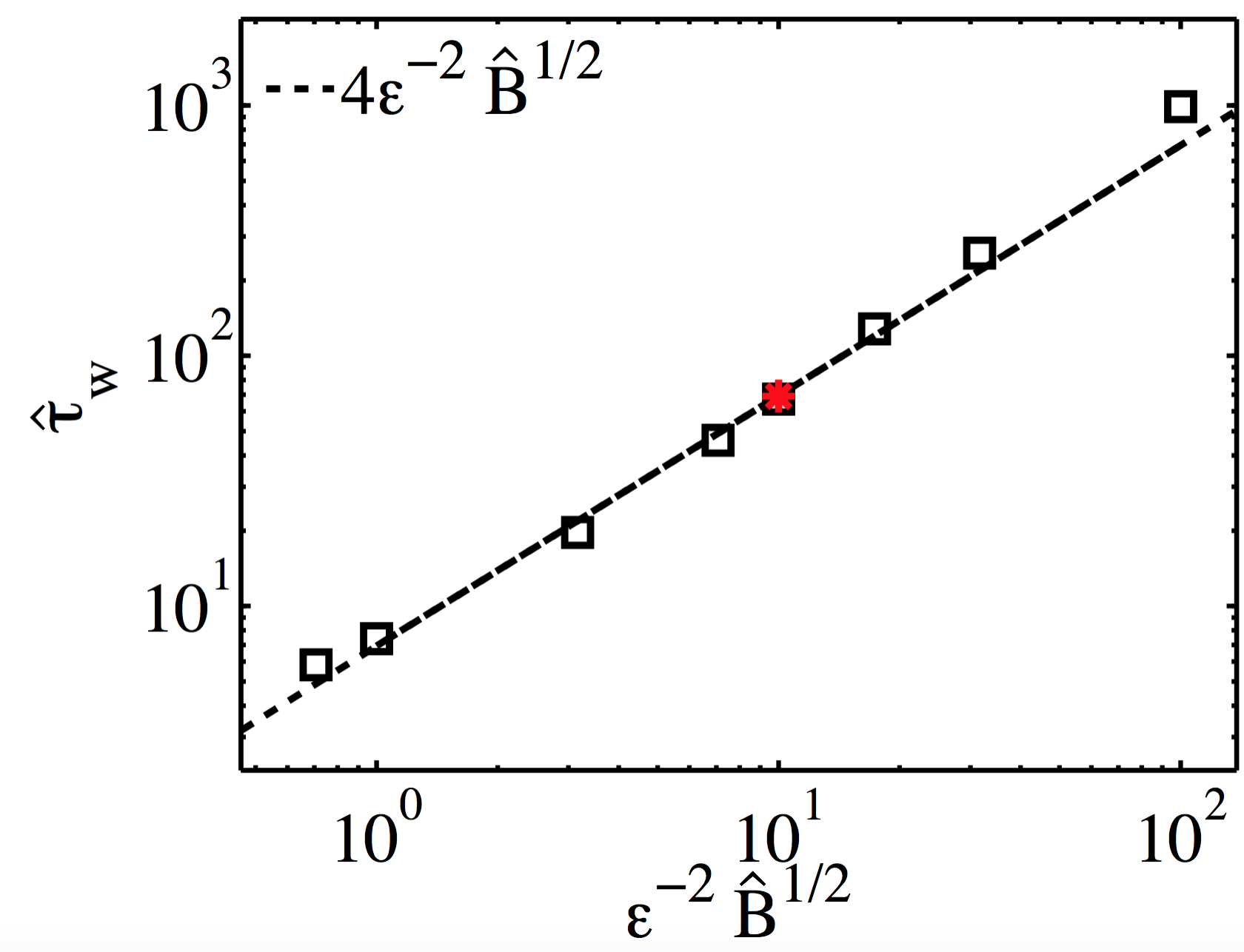}
\caption{Waiting time for onset of adhesion. The scaled waiting time $\hat \tau_w=C(3B/A)^{\frac{1}{2}} =C\epsilon^{-2}(3\hat B)^{\frac{1}{2}}$ for the onset of adhesion as a function the the scaled bending stiffness $\hat B\in[5\times10^{-9}-10^{-4})$, the scaled cut-off height $\hat \sigma\in [0.03-0.12]$ and the aspect ratio is $\epsilon=10^{-2}$, with $C=4$. Square shaped markers represent $\hat B\in[5\times10^{-9}-10^{-4}),~\hat \sigma=0.12$ and the star shaped markers (they are coincident) represent $\hat B=10^{-6}, \hat \sigma= [0.03, 0.06, 0.12]$, the dashed line represents the non-dimensional form of the scaling law in (\ref{eq:waiting}).\label{fig:waiting}}
\end{figure}

\begin{figure}
\centering
\includegraphics[width=0.7\linewidth]{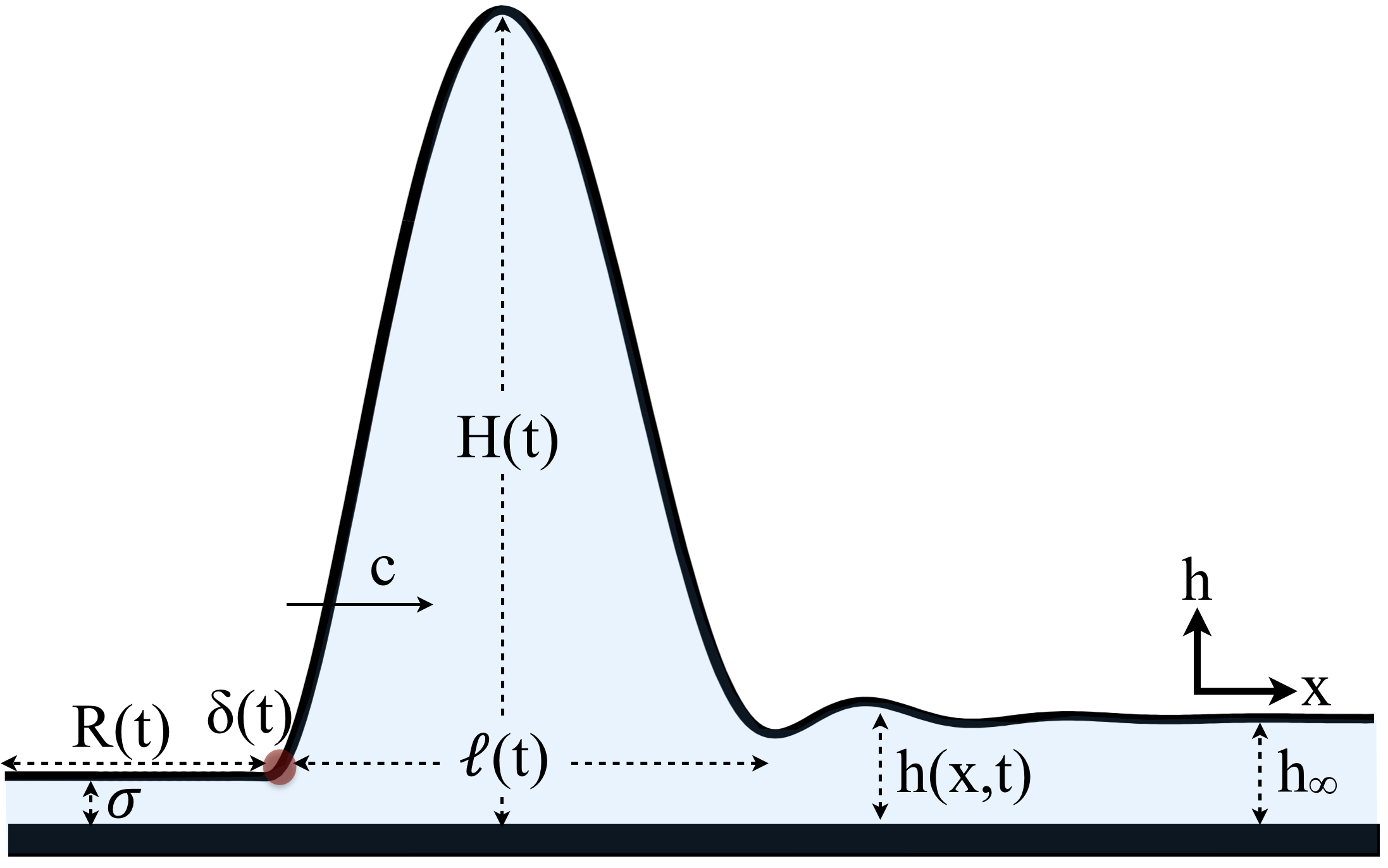}
\caption{A snapshot of the propagating adhesion zone. The fluid filled bump ahead of the adhesion zone moves with a speed $dR(t)/dt=c$. As the adhesion zone grows, it displaces the viscous fluid into the bump making its horizontal extent $\ell(t)$ and height $H(t)$ to grow in time. The circle denotes the small inner region where viscous, elastic and adhesive forces are all comparable in magnitude. Note that the x-axis is compressed by a factor $\approx 1/\epsilon=100$.\label{fig:lengths}}
\end{figure}

\subsection{Waiting time}As the attractive van der Waals pressure $\Phi_h$ brings the surfaces together, fluid is squeezed out. The time for the onset of adhesive contact $\tau_w$ i.e. $h\rightarrow \sigma$ is associated with the time to displace the viscous fluid beneath the adhesion zone. Balancing all three terms in (\ref{tfAdh}) yields the horizontal scale over which pressure and adhesion balance each other when the height of the film is $h_\infty$, i.e. $l_\infty \approx h_{\infty} (B/A)^{\frac{1}{4}}$, and thence  a scaling law for the viscous drainage time
\begin{equation}
\tau_w=C {\mu h_{\infty}^3}(\frac{B}{A^3})^{\frac{1}{2}},
\label{eq:waiting}
\end{equation}
where $C$ is a dimensionless constant. In dimensionless form this reads as $\hat \tau_w=\tau_w/\tau_{\mu}=C\epsilon^{-2}(\hat B)^{\frac{1}{2}}$. Our simulations of [4]-[6] follow this scaling prediction (\ref{eq:waiting}), with $C\approx 4$ see Fig. \ref{fig:waiting}. It is useful to contrast this relation with the capillary (tension) waiting-time for thin fluid film rupture $\approx \frac{\mu \gamma_{LV} h_{\infty}^5}{A^2}$ \cite{carlson:epl}, where $\gamma_{LV}$ is the liquid-vapor surface tension, showing a weaker dependence on both $A$ and $h_{\infty}$.

\subsection{Self-similar shape}

\begin{figure}
\centering
\begin{overpic}[width=0.45\linewidth]{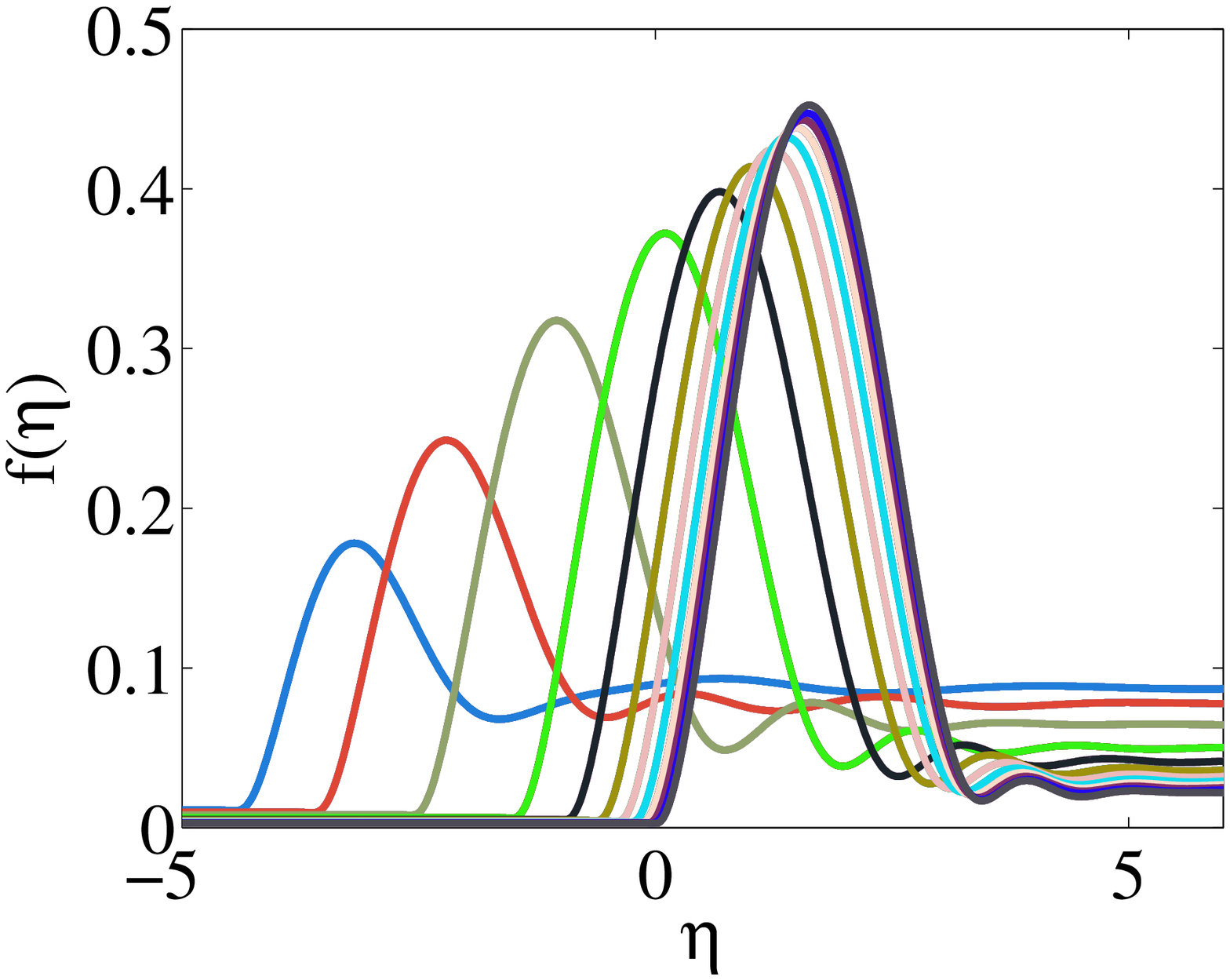}
\put(130,-10){a)}
\end{overpic}
\hspace{0.5cm}
\begin{overpic}[width=0.45\linewidth]{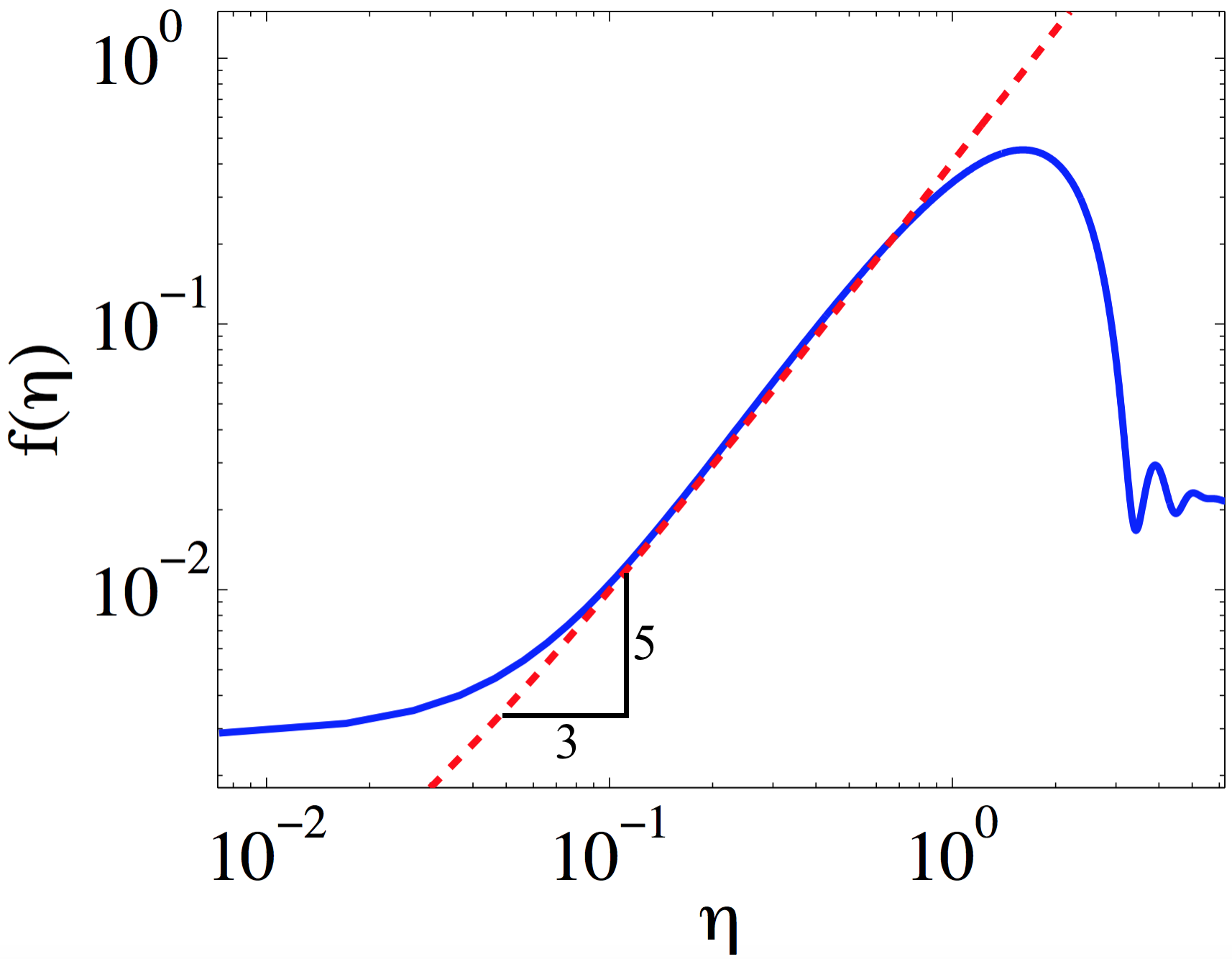}
\put(130,-10){b)}
\end{overpic}
\vspace{0.5cm}
\caption{Evolution of the sheet profile onto a self-similar shape. a) A rescaled version of Fig. \ref{sketch} using the form $f(\eta)=h(x,t)/H(t)$ with $\eta=(x-ct)/\ell(t)$, where $H(t)$ and $\ell(t)$ are given in (\ref{eq:simhl}). The shapes of the sheet collapses onto a universal self-similar shape, here plotted at twelve different points in time starting from left-to-right for $\hat t \in [48-438]$ with $\hat B=5\times 10^{-7}$, $\hat \sigma=0.12$ and the speed $\hat c=0.0133$ is measured numerically. b) One of the shapes of the sheet (solid line) for $\hat t=438$ in logarithmic coordinates, which shows that the slope of $f(\eta)$ is well matched with the analytical prediction (dashed line) in the self-similar regime $f(\eta)\sim \eta^{\frac{5}{3}}$.\label{simshape}}
\end{figure}

\begin{figure}
\centering
\includegraphics[width=0.7\linewidth]{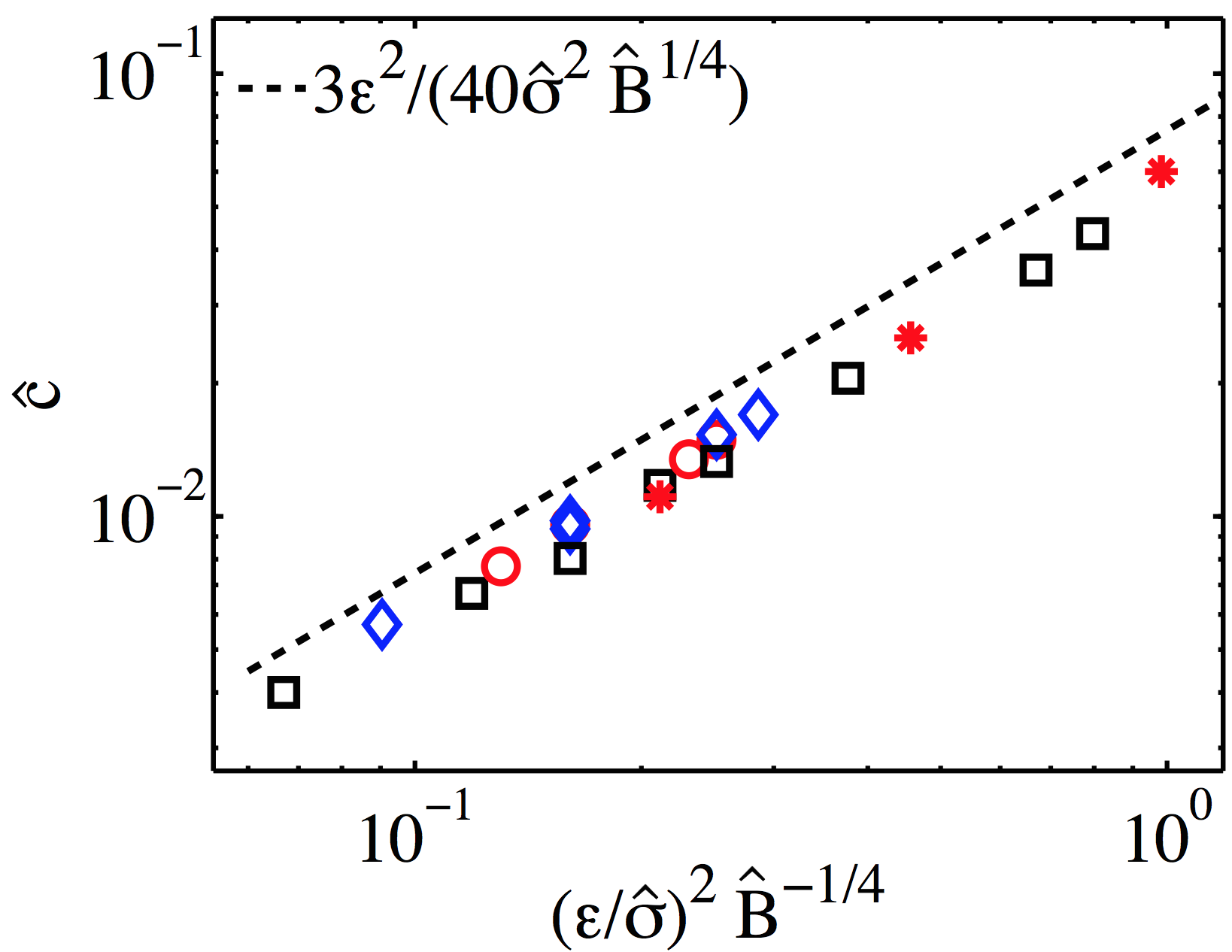}
\caption{Speed of the apparent elastohydrodynamic contact line.  The dimensionless adhesion front speed $\hat c$ as a function of the scaled initial condition corresponding to flat, linear and quadratic initial shape i.e. $\hat h(\hat x,\hat t=0)=[1-0.1\times(1-\tanh(25\hat x)^2,0.5+2\hat x/3,1/2+2\hat x^2/27]$, respectively. The scaled bending stiffness $\hat B\in[5\times10^{-9}-10^{-4}]$ and the scaled cut-off height $\hat \sigma\in [0.03-0.12]$. Squares -- flat, $\hat \sigma=0.12$, $\hat B\in [5\times10^{-9}-10^{-4}]$; circles -- linear $\hat \sigma=0.12$, $\hat B\in [5\times10^{-7}-3\times10^{-5}]$; diamonds -- quadratic, $\hat \sigma=0.12$, $\hat B\in [5\times 10^{-7}-7\times10^{-6}]$; stars -- flat, $\hat \sigma\in[0.03, 0.06, 0.12]$, $\hat B=10^{-6}$. The dashed line represents the scaled form of (\ref{eq:speed}) i.e. $\hat c=3\epsilon^2/(40 \hat \sigma^2 \hat B^{\frac{1}{4}})$.\label{velocity}
}
\end{figure} 

After first contact when $h=\sigma$, the adhesion zone grows by displacing the fluid into a localized bump of height $H(t)$ and width $\ell(t)$, as shown in Fig. \ref{sketch}, and appears to converge to a self-similar traveling form. We also see the formation of an apparent contact line, which we define as the position along the sheet where the curvature is a maximum i.e. $R(t)\equiv \max(h_{xx}(x,t))$.  In the vicinity of the apparent contact line $\max(h_{xx})$ viscous, adhesion and bending forces are all of the same magnitude (see SI). However, as we move away from the apparent contact line the adhesion pressure becomes small and the bump shape is described by elastic bending and fluid conservation, and constrained by the far field boundary conditions. To understand these features, we now turn to a simplified analytical description of the shape and speed of the adhesion zone and also determine the effective boundary conditions at the apparent contact line. 

We start by looking for the asymptotic form of the bump ahead of the adhesion zone, where adhesive effects may be neglected locally. We use a self-similar ansatz seeking a solution to (\ref{tfAdh}) for the film height and the fluid pressure of the form,
\begin{equation}
\begin{split}
&h(x,t)= H(t)f(\eta),~\eta=\frac{x-ct}{\ell(t)}\\
& p(x,t)=P(t)g(\eta).\\ 
\end{split}
\label{eq:simvar}
\end{equation}
Here $(\cdot)' \equiv \partial (\cdot)/\partial \eta$, where $\eta$ is the scaled coordinate frame of the apparent contact line moving at the speed $c$, $f(\eta)$ is the scaled film height, $g(\eta)$ is the scaled pressure, and $H(t)$, $\ell(t)$ and $P(t)$ are three unknown scaling parameters as illustrated in Fig. \ref{fig:lengths}.

In the regime associated with the steadily moving adhesion zone, the characteristic height of the film $H(t)\gg \sigma$ so that the adhesive pressure $A/3H^3(t)f^3(\eta)\ll1$ and the total pressure $p(x,t) \approx B h_{xxxx}  \approx\frac{B}{\ell^4(t)}Hf''''(\eta)$. Furthermore, $h_t=-cf'(\eta)H(t)/\ell(t)+cf(\eta)dH(t)/dt-f'(\eta)H(t)\eta d\ell(t)/dt/\ell(t)\approx -cf'(\eta)H(t)/\ell(t)$, where the first term is dominant at long times \cite{timederiv}. Then, on substituting the form (\ref{eq:simvar}) into (\ref{tfAdh}) we obtain the asymptotic equation
\begin{equation}
-\frac{H^3(t) B f^2(\eta) f'''''(\eta)}{12\mu c \ell^5(t)}=1.
\label{eq:ode}
\end{equation}
Defining the pre-factor $k=H^3(t) B /12\mu c \ell^5(t)$, (\ref{eq:ode}) reads as $-k f^2(\eta) f'''''(\eta)=1$ with a polynomial solution given by 
\begin{equation}
f(\eta)={\eta^{\frac{5}{3}}}, ~g(\eta)= \frac{3\eta^{-\frac{7}{3}}}{7k}, ~k=\frac{BH^3(t)}{12\mu c \ell^5(t)}=\frac{243}{280}.
\label{eq:k}
\end{equation}
Having obtained the form of the solution $f(\eta)$, we now turn to determine the rate of fluid mass swept out by the bump and thence determine its vertical and horizontal scale factors $H(t)$ and $\ell(t)$. For a film with an initial constant height $h(x,t=0)= h_{\infty}$ the volume swept by the bump is given by
\begin{equation}
V = \int^{ct}_{0}h_{\infty} dx=h_{\infty}ct= H(t)\ell(t)\int_{-\infty}^{\infty} f(\eta) d\eta.
\label{eq:mass}
\end{equation}
We assume that the scaled displaced mass is normalized using the condition $\int_{-\infty}^{\infty} f(\eta) d\eta\approx\int_{0}^{1} f(\eta) d\eta=1$, consistent with the numerically measured mass in the bump $\approx 0.85$.
Then, from (\ref{eq:k}) and (\ref{eq:mass}) we find that
\begin{equation}
H(t)\approx\left(\frac{12\mu c^6 k h_{\infty}^5}{B}\right)^{\frac{1}{8}} t^{\frac{5}{8}},~\ell (t)\approx\left(\frac{Bc^2h_{\infty}^3}{12\mu k}\right)^{\frac{1}{8}} t^{\frac{3}{8}},
\label{eq:simhl}
\end{equation}
i.e. the shape of the fluid bump ahead of the adhesion zone is a function of the properties of the elastic sheet, the fluid, and the speed of the adhesion front. In Fig. \ref{simshape}, we corroborate these scaling predictions by showing that our numerical simulations of (\ref{tfAdh}) collapse to a universal rescaled shape. Our numerical observations of the bump ahead of the adhesion zone are consistent with experimental observations \cite{Navarro:2013}. Moreover our similarity solution reveals that in the immediate vicinity of the steadily moving front $h(x,t) \approx H(t) f(\eta)\approx H(t) \left((x-ct)/\ell(t)\right)^{\frac{5}{3}}$ (Fig. \ref{simshape}b), consistent with experimental observations near the adhesion zone \cite{Rieutord2005}. While we have focused  here on the case when the far field corresponds to a flat sheet, we can generalize our analysis to an arbitrary initial condition $h(0,x)\propto x^n$, so that the volume swept up by the bump depends on $n$ via the shape of the sheet in the far field (see SI). 

\subsection{Adhesion speed}
Having characterized the self-similar shape of the traveling bump, we now determine the speed $c$ of the apparent contact line. {In a traveling wave frame moving with a speed $c$, our numerical simulations show that the elastic sheet squeezes the liquid ahead, forming a bump that grows with time as it sweeps up the fluid ahead. Once the contact zone reaches a steady state, we expect that no energy is expended in local bending of the sheet, since the size of the contact zone $l_c$  is invariant. Thus, the speed of the apparent contact line should be determined by the balance between the driving adhesive power and the viscous dissipation in the contact zone. 

At a scaling level, the adhesive power per unit width $\Pi_a$ scales as the product of the speed and the energy per unit area of adhesion, i.e. $\Pi_a \sim c \Phi(\sigma) \sim c\gamma \sim cA/\sigma^2$, while the viscous power per unit width $\Pi_d$ scales as $\mu (\nabla U)^2 \Omega $, where $\nabla U \sim c/\sigma$, and the volume per unit width $\Omega \sim\sigma l_c$, so that $\Pi_d \sim \mu c^2 l_c/\sigma$. Balancing the adhesive pover with the dissipation i.e. $\Pi_a \sim \Pi_d$ yields $c \sim A^{\frac{5}{4}}/\sigma^2 \mu B^{\frac{1}{4}}$, which we may rewrite as $c \sim \gamma A ^{\frac{1}{4}}/\sigma^2 \mu B^{\frac{1}{4}}$, an expression very similar to that given in \cite{Rieutord2005}, although we note that by accounting for microscopic adhesion, we do not need to postulate the existence of an apparent slip length used in \cite{Rieutord2005}.

Going beyond this simple scaling analysis, we write the exact balance between the dissipation rate and the adhesion and bending power in the steady state problem as   \cite{Rieutord2005}
\begin{equation}
\int_{0}^{\infty}-\frac{h^3p_x^2}{12\mu} dx=\frac{d}{dt}\left[\int_{0}^{\infty} \left( \frac{B}{2}h_{xx}^2+\Phi(h) \right)dx\right].
\label{eq:energy}
\end{equation}
To understand the relative magnitudes of the different terms, we introduce the similarity variables (\ref{eq:simvar},\ref{eq:simhl}) into (\ref{eq:energy}). In the contact zone, the elastic bending pressure and the attractive van der Waals pressure balance each other so that $\frac{A}{3h^3} = Bh_{xxxx}$, i.e. $\frac{A}{3H^3(t)f^3(\eta)} = \frac{BH(t)f''''}{\ell^4(t)}$. As we approach the apparent contact line, where the sheet is statically adherent, with $\eta\rightarrow \delta(t)$ and $f=\eta^{\frac{5}{3}}$, we find that
\begin{equation}
\delta(t) \approx \left(\frac{A}{3B}\right)^{\frac{3}{8}}\left(\frac{\ell (t)}{H(t)}\right)^{\frac{3}{2}}=\left(\frac{A}{36\mu c^2 k h_{\infty}}\right)^{\frac{3}{8}}t^{-\frac{3}{8}}.
\label{eq:delta}
\end{equation}

\begin{figure}
\includegraphics[width=0.70\linewidth]{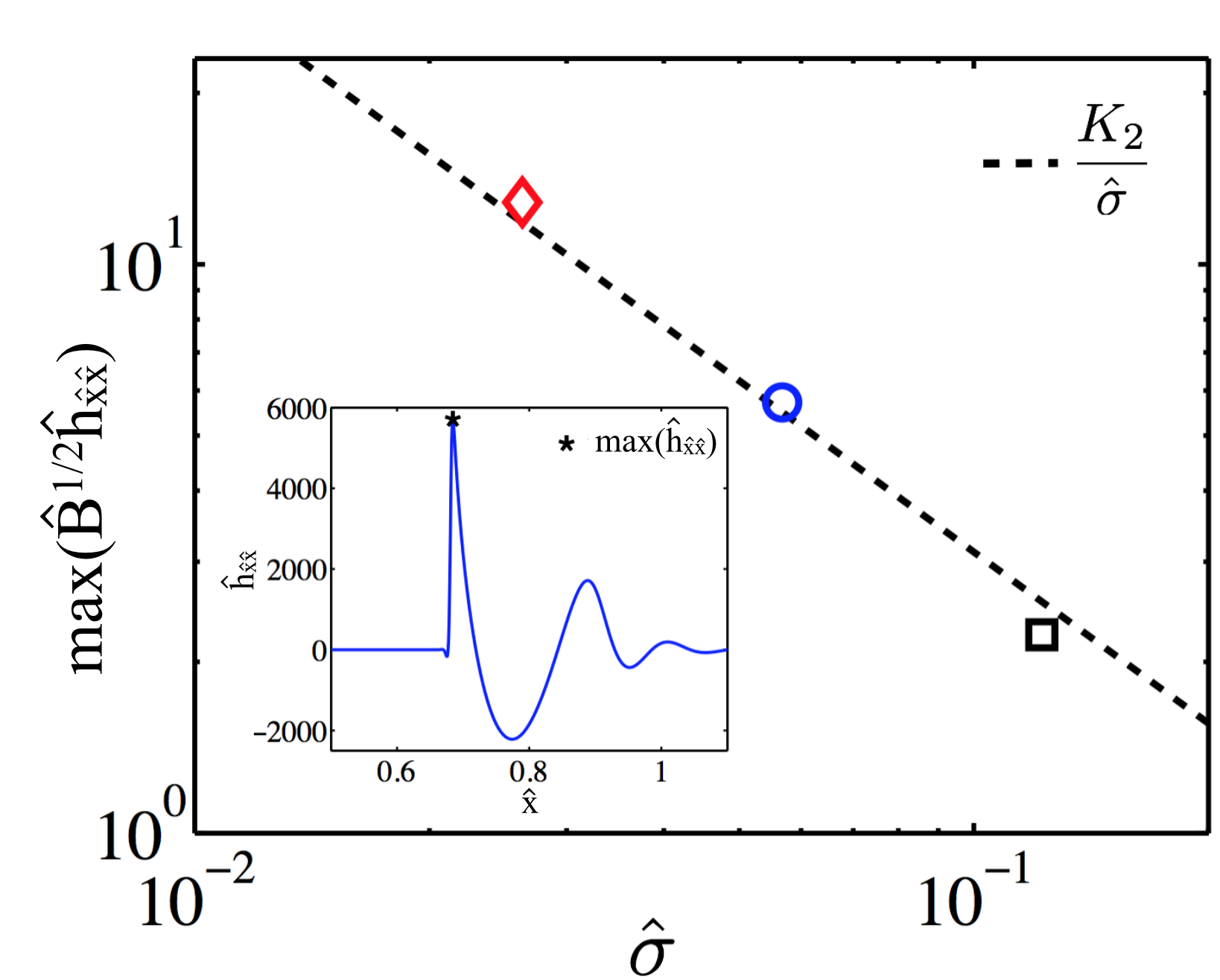}
\caption{Characterizing the elastohydrodynamic contact zone. The scaled maximum curvature in the contact zone $\max( \hat B^{\frac{1}{2}}{\hat h_{\hat x\hat x})}$ with the scaled bending stiffness $\hat B=10^{-6}$ and three different scaled minimum heights $\hat \sigma=[0.03, 0.06, 0.12]$. The dashed line is the scaled form of the relation (\ref{eq:BCel}) and illustrates the singular nature of the curvature as $\hat \sigma \rightarrow 0$. Inset: The curvature of the sheet $\hat h_{\hat x \hat x}$ with $\hat \sigma=0.06$, where the star shaped marker defines the position where the curvature is a maximum.
\label{fig:hxx}}
\end{figure}

First, we note that the rate of change of elastic bending energy reads as
\begin{equation}
\begin{split}
&\frac{d}{dt} \int_{-\infty}^{\infty} \frac{B}{2}h_{xx}^2 dx\approx\\
&\left({B^3(12\mu k)^5h_{\infty}c^{6}}\right)^{\frac{1}{8}} t^{-\frac{7}{8}}\int_{\delta(t)}^{1} \eta ^{-\frac{2}{3}}d\eta.
\end{split}
\label{eq:simhxx}
\end{equation}
Inserting (\ref{eq:simhl},\ref{eq:delta}) in (\ref{eq:simhxx}) shows that at its lower limit the integral scales as $ t^{-\frac{7}{8}}\delta(t)^{\frac{1}{3}}\sim t^{-1}$ and at the upper limit the integral scales as $t^{-\frac{7}{8}}$. Thus, although the integrand diverges in the vicinity of the contact line, the integral itself becomes vanishingly small at long times. Physically this is consistent with the idea that bending energy is neither stored nor dissipated as the front has an invariant shape that persists with time. 

Next, we note that the viscous dissipation rate reads as
\begin{equation}
\int_{0}^{\infty} -\frac{h^3p_x^2}{12\mu}dx\approx-\frac{H^5(t)B^2}{12\mu l^9(t)k^2}\int_{\delta(t)}^{1}\eta^{-\frac{5}{3}} d\eta.
\label{eq:simenergy}
\end{equation}
In the neighborhood of the contact line, the integral remains bounded due to the presence of the cut-off length $\delta(t)$. Indeed, in the asymptotic regime $t\rightarrow \infty$ the value of the integral at its upper limit scales as $\propto t^{-\frac{2}{8}}$ and thus becomes vanishingly small. The value of the integral is instead dominated by the lower limit. Inserting $\delta(t)$ from (\ref{eq:delta}) in (\ref{eq:simenergy}) yields an estimate for the viscous dissipation as $\sim \mu c^2 (B/A)^{\frac{1}{4}}$, which is constant for a front traveling with a constant speed, consistent with our simulations (see SI).

Finally, we note that the adhesive power is dominated by the contribution from the maximum of the adhesion potential i.e.
\begin{equation}
\frac{d}{dt} \int_{0}^{\infty} \Phi(h) dx\approx \frac{d}{dt} \int_{0}^{R(t)} \max \left(\Phi(h)\right) dx\approx \frac{Ac}{8\sigma^2}.
\label{eq:rocad}
\end{equation}
Balancing the rate of change of adhesion energy (\ref{eq:rocad}) with the viscous dissipation rate (\ref{eq:simenergy}) yields a scaling law for the adhesion front speed 
\begin{equation}
c\approx \frac{A^{\frac{5}{4}}}{160 \sigma^2 \mu B^{\frac{1}{4}}},
\label{eq:speed}
\end{equation}
which in dimensionless form reads as $\hat c=3\epsilon^2/(40\hat\sigma^2 \hat B^{\frac{1}{4}})$, consistent with our earlier simple scaling estimate. Since $A/8\sigma^2 = \gamma$, the interfacial tension, we see that the speed is independent of the microscopic length scale $\sigma$, but does depend on the adhesion strength $A$, consistent with earlier results \cite{Rieutord2005}.

In Fig. \ref{velocity}, we show the speed $c$ of the adhesion front obtained by solving (\ref{tfAdh}) for a range of parameter values of $\hat B\in[5\times10^{-9}-10^{-4}]$, initial far-field shapes determined by $n$, cut-off heights $\sigma \in [0.03-0.12]$ for the aspect ratio $\epsilon=10^{-2}$, and see that the simulations corroborate (\ref{eq:speed}). Our analytical prediction also favorably compares with experiments \cite{Bengtsson1996,Rieutord2005,Navarro:2013}. By using their material parameters in (\ref{eq:speed}) i. e. $\gamma=0.12$ N/m, $\mu=2\times 10^{-5}$ Pa$\cdot$s, $B=4$ Nm \cite{Rieutord2005,Navarro:2013} and $A=3\times10^{-19}$ Nm \cite{Israelachvili2011}, gives a theoretical wafer bonding speed $c \approx 0.5$ cm/s consistent with the experimental observations that yield $c \approx 0.2 - 2 cm/s$\cite{Bengtsson1996,Rieutord2005,Navarro:2013}. 


\subsection{Boundary conditions at the apparent elastohydrodynamic contact line}
We now turn to address the question of the singular nature of the shape of the adherent sheet in the adhesion zone and by using (\ref{eq:simvar}, \ref{eq:k}, \ref{eq:simhl},\ref{eq:delta}) we can evaluate the similarity form for $h_x\approx H(t)f'(\eta)/\ell(t) \approx H(t)\eta^{\frac{2}{3}} /\ell(t)\approx H(t)\delta^{\frac{2}{3}}/\ell(t) \approx (\frac{A}{B})^{\frac{1}{4}}$. Thus, our analysis suggests that the apparent contact angle $h_x$ of the elastic film approaches a constant value in the contact zone, similar to the Young-Laplace contact condition for a static fluid-fluid contact line on a solid substrate \cite{degennes85, Bonn2009, Snoeijer2013}. This implies that the curvature of the sheet diverges in this zone. Of course, this divergence is suppressed in the adhesion zone over a size $l_c\approx \sigma (B/A)^{\frac{1}{4}}$  determined by the balance between the bending and the attractive van der Waals pressure. Indeed, in the contact zone $h_x\approx \sigma/l_c\approx (A/B)^{\frac{1}{4}}$ and $h_{xx}\approx \sigma/l_c^2\approx (A/B)^{\frac{1}{2}}\sigma^{-1}$. In Fig. \ref{fig:hxx}, we show that the scaled maximum curvature in the contact zone ${\rm max}(h_{xx})$ is in agreement with both this scaling law and the similarity solution. 


Finally, we determine how flow and elastic deformation processes together yield the effective boundary conditions associated with mesoscopic deformations of the elastic sheet, viewed from the perspective of the outer problem where the microscopic adhesion is relatively unimportant. Approaching the dynamic apparent elastic contact line $R(t)$ from the right side, we note that outside the contact zone, the van der Waals adhesion pressure is negligible, so that the equation of motion \ref{tfAdh} simplifies to $12 \mu h_t-(Bh^3h_{xxxxx})_x=0$. Looking for a traveling wave solution of the form $h(x,t) = g(x-ct)$ then simplifies the equation to the form $-cg'-(Bg^3g''''')'=0$, where $(\cdot)'=d(\cdot)/d(x-ct)$ which has a similarity solution of the form $g \sim (x-ct)^{5/3}$ consistent with \cite{Rieutord2005}, and our earlier similarity analysis. We note that as we approach the contact line, $x-ct \rightarrow 0$, $g \rightarrow 0, g' \rightarrow 0$, but the curvature $g'' \rightarrow \infty$. However, this divergence is ameliorated by the presence of the inner scale associated with the contact zone where microscopic adhesion, bending and fluid flow are in balance. This leads us to the conclusion that the effective boundary conditions for the outer problem at the contact line $R(t)$, i.e. ignoring the details of the contact zone, are given as
\begin{equation}
h(R)=0, ~~h_{x}(R) = 0,~~ c = \frac{dR}{dt}=K_1\frac{A^{\frac{5}{4}}}{160\sigma^2 \mu B^{\frac{1}{4}}}
\label{eq:BCel}
\end{equation} 
where $K_1 =0.8\pm0.1$ is a dimensionless constant evaluated numerically. In dimensionless form these read as; $\hat h(\hat R)=0,\hat h_{\hat x}(\hat R)=0,\hat c= 3K_3\epsilon^2/(40\hat\sigma^2 \hat B^{\frac{1}{4}})$. It is useful to contrast these conditions with those for a static adhered sheet, where at the elastic contact line \cite{Obreimoff1930} i.e. $h(R)=h_{x}(R)=0, h_{xx}(R)=(\frac{2\gamma}{B})^{\frac{1}{2}}=(\frac{A}{4\sigma^2B})^{\frac{1}{2}}$, i.e. the static condition for the curvature at the contact line is replaced by a condition for the speed of the apparent contact line.  In the limit $\sigma=0,A=0$, the boundary conditions in (\ref{eq:BCel}) read $h=h_x=c=0$, and the contact line becomes stationary \cite{Lister2013PRL,fitton:2004},  consistent with the classical boundary conditions for a non-adherent elastic sheet in contact with a solid.

These results might also be contrasted with the analogous problem of describing static and dynamic contact lines in interfacial hydrodynamics where a static contact line has a constant contact angle, whereas a dynamic contact line has a contact angle that is a function of its speed \cite{degennes85,Bonn2009,Snoeijer2013}. 

\section{Conclusions}
Our theoretical study of viscously limited elastohydrodynamics of adherent sheets captures the entire process of dynamical adhesion, from the short-time onset of adhesion to the long time dynamics associated with a steadily propagating adhesion front.  A simple mathematical model provides a compact formulation that naturally couples the microscopic physics at the apparent elastic contact line and the macroscopic physics associated with elastic deformation and fluid flow, complementing earlier preliminary analyses of the problem. Numerical simulations reveal different regions of the sheet; an adherent zone $x<R(t)$ at constant height $\sigma$, an inner contact zone where viscous flow, elastic bending and microscale adhesion are all equally important, and an outer region with $h\gg\sigma$ where the sheet is described by viscous flow and elastic bending. An asymptotic similarity analysis of the governing partial differential equation allows us to describe the outer zone consistent with the solutions of the governing partial differential equation, and leads to a self-similar shape of the elastic sheet and its propagation speed. In addition, we have derived the effective boundary conditions for the dynamic apparent elastic contact line, which highlights its singular nature and distinguishes it from analogous contact line conditions for a static sheet. 


Just as high resolution Total Internal Reflection Fluorescence microscopy allowed for a more detailed view of the dynamics of liquid contact lines \cite{Snoeijer2013}, we hope that our theoretical study might engender further investigation of the nature of the elastohydrodynamic contact line that arises in a wide range of problems where elastic interface adheres to solid substrates in fluid environments.

\section*{Acknowledgments and Contributions}
We thank the Harvard-MRSEC DMR-1420570, and the MacArthur Foundation (LM) for support.  AC, SM and LM conceived of the mathematical model, did the research and drafted the manuscript. AC did most of the simulations. All authors gave final approval for publication. The authors have no competing interests.

\renewcommand{\figurename}{Fig. A}
\renewcommand{\theequation}{A\arabic{equation}}
        \setcounter{figure}{0}
        \setcounter{equation}{0}

\clearpage
\section{Appendix}

\subsection{Static sheet}
Long wavelength deflections in a static sheet yield the equation of equilibrium for its shape that is given by
\begin{equation}
p=Bh_{xxxx}- \Phi_h.
\label{eq:SIp}
\end{equation}
The accompanying boundary conditions are $h(0)=h_x(0)=0$ and at the support, $h(L)=h_{\infty},~h_{xx}(L)=0$, see Fig. A\ref{static}. We note that  (\ref{eq:SIp}) follows as the Euler-Lagrange equation associated with the energy functional $ \int_R^L [B h_{xx}^2/2 + \Phi (h) ] dx$. Taking the variation of this energy with respect to the location of the the contact line at $x=R$, the principle of virtual work yields the contact line conditions: 
\begin{equation}
h(R)= \sigma,  h_x(R)=0, \\
B h_{xx}^2(R) = 2\Phi(\sigma),
\end{equation}
where the last condition can be rewritten as $h_{xx}(R)= (\frac{A}{4\sigma^2B})^{\frac{1}{2}}=(2\gamma)^{\frac{1}{2}}$. Similarly, these boundary conditions can be obtained by multiplying [1] by $h_x$ and integrating by parts.  
\begin{figure}
\centering
\includegraphics[width=0.7\linewidth]{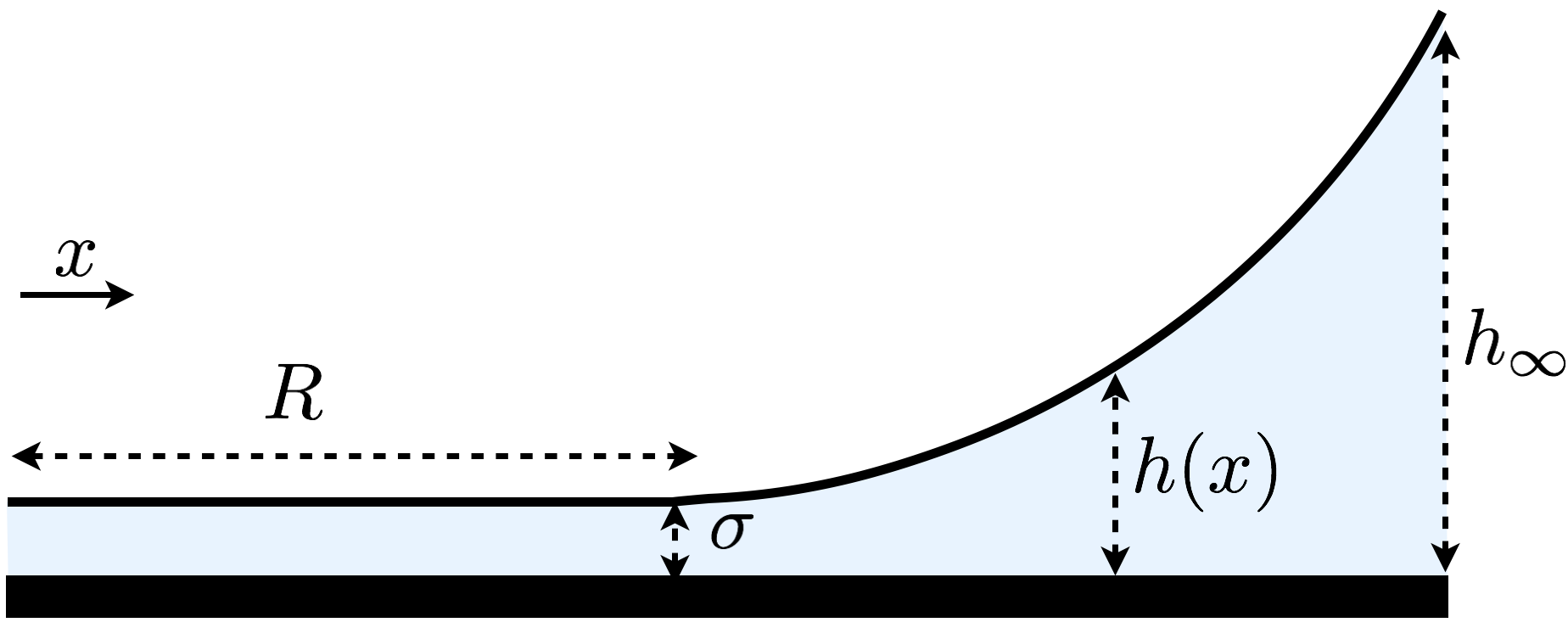}
\caption{Schematic. A static elastic sheet that is adhered to a solid substrate at is left edge at an equilibrium height $h=\sigma$, while the right end is supported at a height $h_{\infty}$. Note that the x-axis is compressed by a factor $\approx 1/\epsilon=100$.\label{static}}
\end{figure}

\subsection{Role of initial shape of the sheet on its dynamics}
Different initial shapes affect the dynamic shape of the sheet during adhesion. First we recall the similarity ansatz and the polynomial solution to the self-similar equation in the regime where $A/h^3\ll1$, i.e. where adhesion can be neglected,
\begin{equation}
\begin{split}
&h(x,t)= H(t)f(\eta),~\eta=\frac{x-ct}{\ell(t)},\\
&f(\eta)={\eta^{\frac{5}{3}}}, ~g(\eta)= \frac{3\eta^{\frac{-7}{3}}}{7k}, ~k=\frac{BH^3(t)}{12\mu c \ell^5(t)}=\frac{243}{280}.
\end{split}
\label{eq:SIk}
\end{equation}
We account for a generic initial shape of the form $h(x,t=0)\approx eh_{\infty}(x/L)^n$ corresponding to flat, linear and quadratic profiles and $e$ is a constant pre-factor. The volume swept by the bump is given by
\begin{equation}
V(n)\approx\int^{ct}_{0}e h_{\infty}(\frac{x}{L})^n dx\approx\frac{L h_{\infty}e (ct/L)^{n+1}}{n+1}=H(t)\ell (t) \int_{-\infty}^{\infty} f(\eta) d\eta.
\label{eq:SImass}
\end{equation}
We normalize the scaled displaced mass using the condition $\int_{-\infty}^{\infty} f(\eta) d\eta\approx \int_{0}^{1} f(\eta) d\eta=1$ and the numerical simulations indicate that this is a fairly good approximation, as seen in Fig. 5 where the numerical area of the bump is $\approx 0.85$. 

Then, substituting (\ref{eq:SIk}) in (\ref{eq:SImass}) we find that
\begin{equation}
H(t)\approx\left(\frac{12\mu c k V^5(n)}{B}\right)^{\frac{1}{8}},~\ell (t)\approx\left(\frac{BV^3(n)}{12\mu c k}\right)^{\frac{1}{8}},
\label{eq:SIsimhl}
\end{equation}
i.e. the dynamic shape of the elastic sheet is a function of the fluid, the adhesion speed $c$ and the initial film shape through $V(n)$. We see that the initial condition determines the dynamics of the height and width of the bump, corresponding to; constant height [$n=0$, $H(t)\propto t^{\frac{5}{8}}$, $\ell(t)\propto t^{\frac{3}{8}}$], linear height, [$n=1$, $H(t) \propto t^{\frac{3}{4}}$, $\ell(t)\propto t^{\frac{5}{4}}$] and quadratic height [$n=2$, $H(t) \propto t^{\frac{9}{8}}$, $\ell(t)\propto t^{\frac{15}{8}}$] but does not affect the velocity of the adhesion front, see Fig. 6. To test if the self-similar ansatz holds we have scaled three membrane shapes at different points in time and with very different mass accumulated in the bump, see Fig. A\ref{SI:shape}a. Although the dynamics of the bump is a function of the initial condition, it appears to converge to a universal self-similar shape. Note that a pre-factor [0.8, 0.66] for both $H(t)$ and $\ell(t)$ is adjusted for the scaling of $n=[1,2]$, respectively. Determining this scaling factor analytically requires us to match the different regions which is a function of the initial condition that appears to have a small influence on the volume integral $\int_{-\infty}^{\infty} f(\eta) d\eta$.

\subsection{Dissipation}
Since the apparent elastic contact line is found to propagate with a constant velocity, we anticipate that the viscous dissipation rate in the thin fluid film is constant. To verify this, we extract the viscous dissipation rate, given by equation [16] in the main text, from one of our numerical simulations as shown in Fig. A\ref{fig:SIdissipation}; it approaches a constant value after the initial transients die out. 

\begin{figure}
\centering
\begin{overpic}[width=0.425\linewidth]{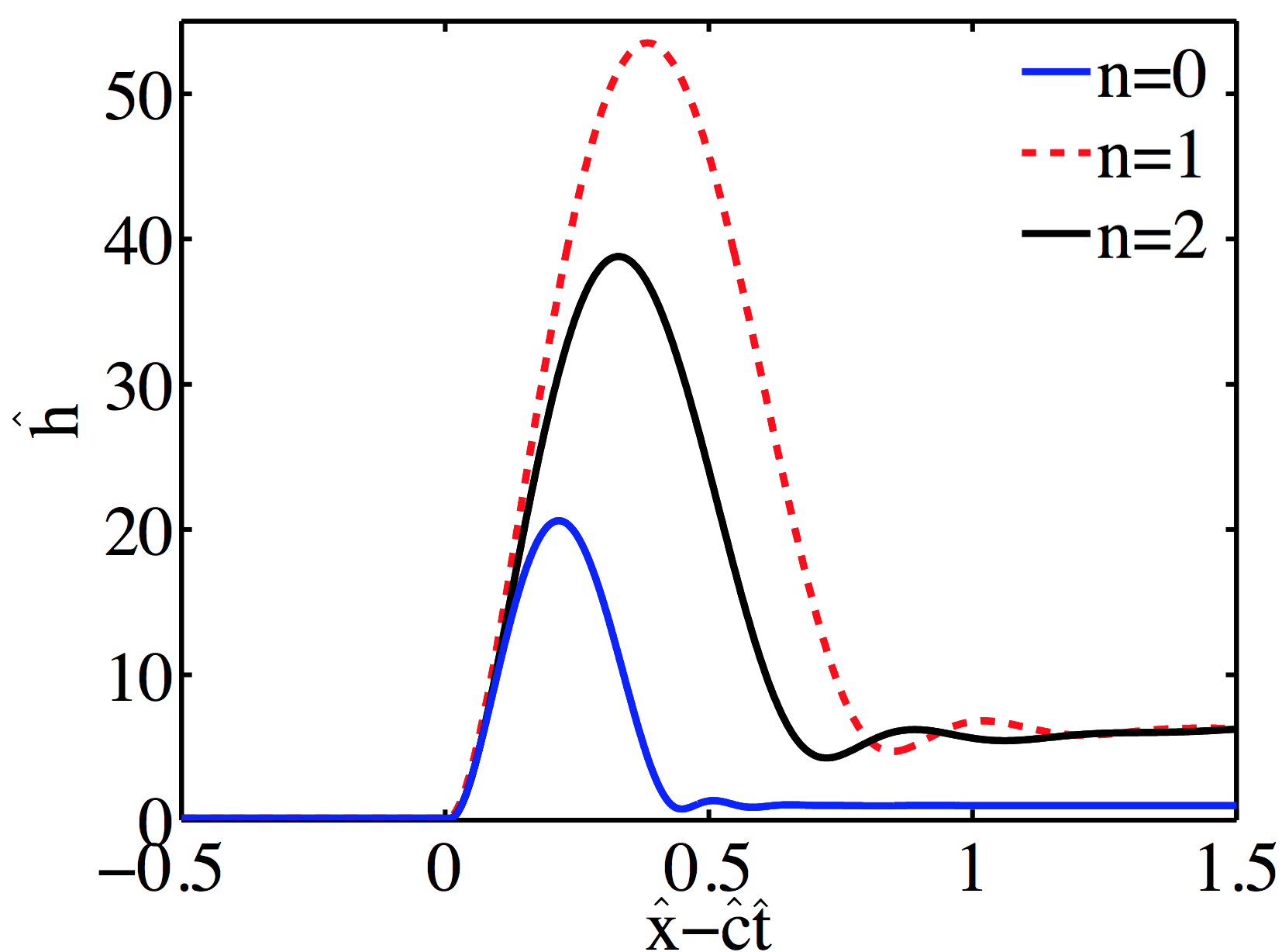}
\put(130,-10){a)}
\end{overpic}
\hspace{2cm}
\begin{overpic}[width=0.425\linewidth]{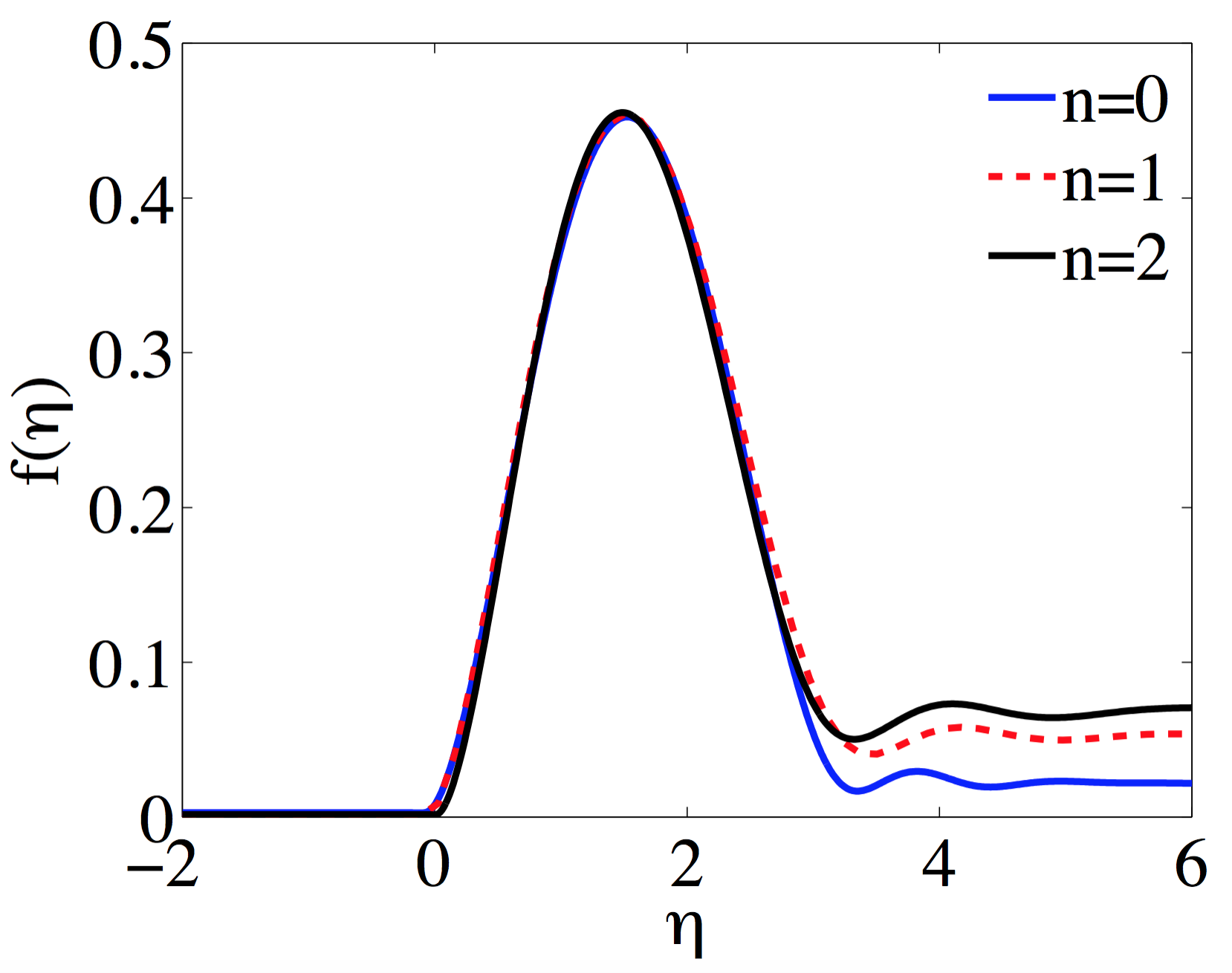}
\put(130,-10){b)}
\end{overpic}
\vspace{0.5cm}
\caption{Self-similar shape of the adhesion front. a) The scaled shape of the sheet $\hat h$ for three different initial conditions $n=[0, 1, 2]$ for $\hat t=[438, 458, 478]$, respectively. $\hat h(\hat x,\hat t=0)$ is; constant height $n=0$, $\hat h(\hat x,\hat t=0)=1-0.1\times(1-\tanh(25\hat x)^2$, linear height $n=1$, $\hat h(\hat x,\hat t=0)=1/2+2\hat x/3$, quadratic height $n=2$, $\hat h(\hat x,\hat t=0)=1/2+2\hat x^2/27$. b) Rescaled height and x-coordinate according to (\ref{eq:SIsimhl}), i.e. $f(\eta)=h/H(t)$, $\eta=(x-ct)/\ell(t)$ gives the self-similar shape of the bump.We need different pre-factors to collapse the shapes i.e $n=1,$ $0.8\times[H(t,n=1),\ell(t,n=1)]$ and$n=2$, $0.66\times[H(t,n=2),\ell(t,n=2)]$, however both pre-factors are $O(1)$ and the influence on the detailed mass balance $\int_{-\infty}^{\infty} f(\eta) d\eta$ may come from the far-field shape of the sheet that requires matching to the self-similar region.\label{SI:shape}}
\end{figure}

\begin{figure}
\centering
\includegraphics[width=0.60\linewidth]{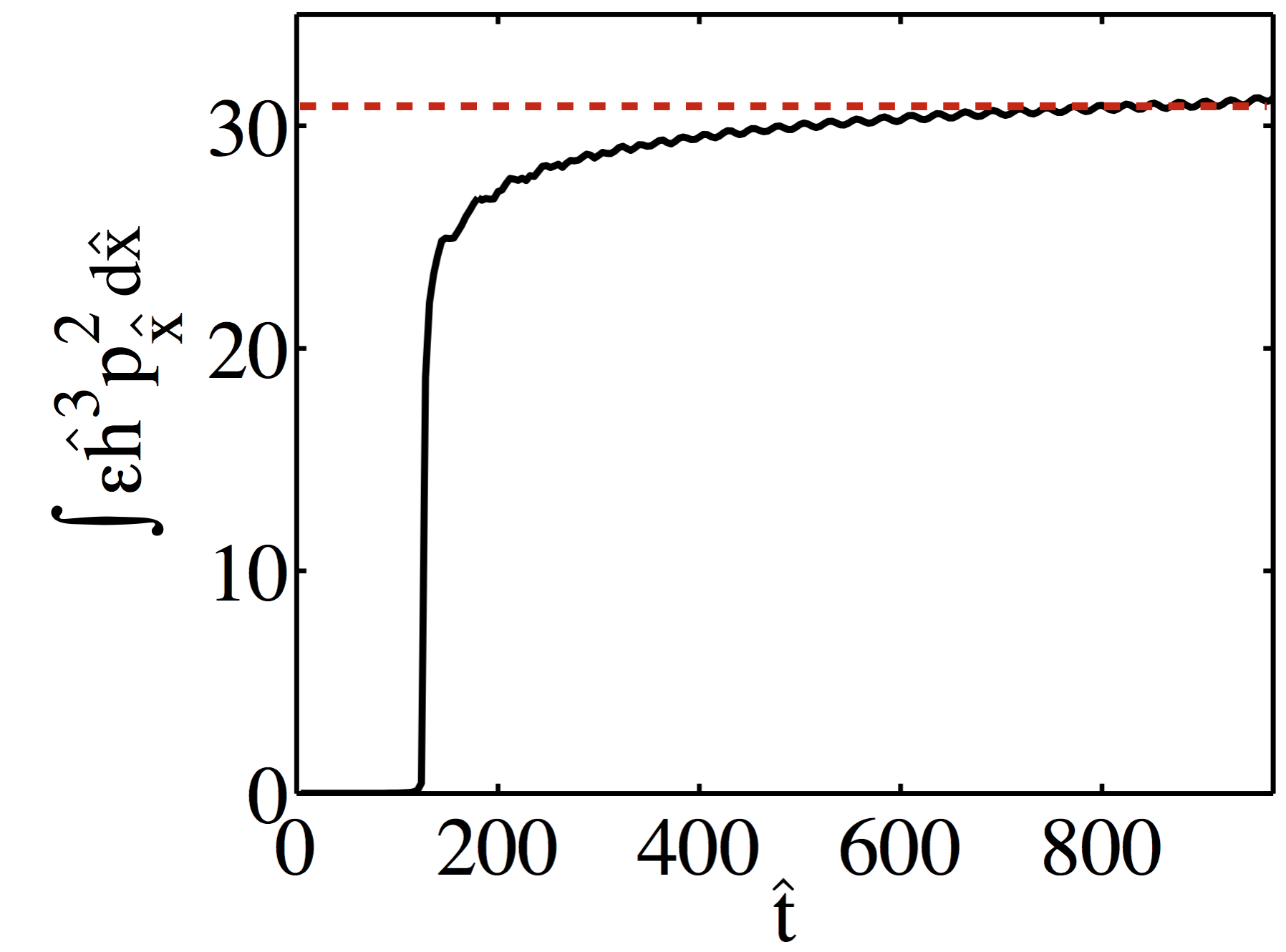}
\caption{The solid line shows the non-dimensional viscous dissipation as a function of time for the parameters; $\hat B=3\times10^{-6}$, $n=0$, $e(0)=1$ $\hat \sigma=0.12$. At short times, as the adhesion zone is formed, viscous dissipation rapidly increases. At long times the adhesion front propagates with a constant speed $\hat c$ and the viscous dissipation approaches a constant value. The dashed line is intended as a guide to the eye.
\label{fig:SIdissipation}}  
\end{figure}


\begin{thebibliography}{100}
 
\bibitem{Bell:1978}
Bell G. I. (1978) {Models for the specific adhesion of cells to cells.} {\em Science} {\bf 200}, 618--627.

\bibitem{Giannone:2007} 
Bongrand, P. (1988) {Physical basis for cell adhesion.} {\em CRC Press}. 

\bibitem{Chaudhury2005}
Chaudhury M. and Pocius A. V. (Editors). (2002) {\em Adhesion Science and Engineering: Surfaces, Chemistry and Applications.} (Elsevier Science B.V., The Netherlands).

\bibitem{Tong}
Tong Q.-Y. and  G\"osele U.  (1999) {Semiconductor wafer bonding: science and technology}. {\em Wiley, New York}

\bibitem{Bengtsson1996}
Bengtsson, S., Ljungberg, K. and Vedde, J. (1996) {The influence of wafer dimensions on the contact wave velocity in silicon wafer bonding}. {\em Journal of Applied Physics Letters} {\bf 69}, 3381.

\bibitem{Turner:2004}
Turner, K.T., Thouless, M.D. and Spearing, S.M. (2004) {Mechanics of wafer bonding: effect of clamping}. {\em Journal of Applied Physics} {\bf 95}, 349--355.

\bibitem{weihua:2000}
Weihua, H., Jinzhong, Y. and Qiming W. (2000) {Modeling the dynamics of Si wafer bonding during annealing}. {\em Journal of Applied Physics} {\bf 88}, 12729--12734.

\bibitem{Rieutord2005}
Rieutord, F., Bataillou, B. and Moriceau, H. (2005) {Dynamics of a bonding front}. {\em Physical Review Letters} {\bf 94}, 236101.

\bibitem{Radisson:2013}
Radisson, D., Fournel, F.  and Charlaix, E.  (2013) {Modelling of the direct bonding wave.} {\em Microsystem Technology} {\bf 103}, DOI 10.1007/s00542-015-2445-3.

\bibitem{Navarro:2013}
Navarro, E., Br\'echet, Y., Moreau, R., Pardoen, T., Raskin, J.-P., Barthelemy, A. and Radu, I.  (2013) {Direct silicon bonding dynamics: a coupled fluid/structure analysis}. {\em Applied Physics Letters} {\bf 103}, 034104.

\bibitem{Bell:1984} 
Evans, E.A. (1985) {Cell adhesion: competition between nonspecific repulsion and specific bonding.} {\em Biophysics Journal} {\bf 45}, 1051--1064.

\bibitem{Cantat:1998}
Cantat, I.  and Misbah, C. (1998) {Dynamics and Similarity Laws for Adhering Vesicles in Haptotaxis.} {\em Physical Review Letters} {\bf 57}, 235.

\bibitem{Chaudhury2013EPL}
Longley, J. E., Mahadevan, L. and Chaudhury M. K. (2013) {How a blister heals.} {\em Europhysics Letters} {\bf 104}, 46002.

\bibitem{Mani:2012} 
Mani M., Gopinath A. and Mahadevan L. (2012) {How things get stuck: kinetics and elastohydrodynamics of soft adhesion}. {\em Physical Review Letters} {\bf 108}, 226104.

\bibitem{Hosoi:2004} Hosoi A. E. and Mahadevan L. (2007) {Peeling, healing and bursting in a lubricated elastic sheet}. {\em Physical Review Letters} {\bf 93}, 137802-1--4.

\bibitem{Lister2013PRL}
Lister, J.R., Peng, G.G., and Neufeld, J.A. (2013) {Viscous control of peeling an elastic sheet by bending and pulling}. {\em Physical Review Letters} {\bf 111}, 154501.

\bibitem{Leong}
Leong, F. Y. and Chiam, K.-H. (2010) {Adhesive dynamics of lubricated films}. {\em Physical Review E} {\bf 81}, 041923.

\bibitem{Pihler:2012}
Pihler-Puzovic, D., Illien,P., Heil, M. and Juel, A. (2012) {Suppression of complex fingerlike patterns at the interface between air and a viscous fluid by elastic membranes}. {\em Physical Review Letters} {\bf 108}, 074502.

\bibitem{michaut}
Michaut, C.  (2011){Dynamics of magmatic intrusions in the upper crust: Theory and applications to laccoliths on Earth and the Moon}. {\em Journal of Geophysical Research}  {\bf 116}, B05205.

\bibitem{Bruyn:2015}
Hewitt, I. J., Balmforth, N. J. and De Bruyn, J. R. (2015) {Elastic-plated gravity currents}. {\em European Journal of Applied Mathematics} {\bf 26}, 1.

\bibitem{fitton:2004}
Fitton, J. C. and King, J. R. (2004) {Moving-boundary and fixed-domain problems for a sixth-order thin-film equation}. {\em European Journal of Applied Mathematics} {\bf 15}, 713.




\bibitem{Rieutord:2014}
Rieutord, F., Rauer, C. and Moriceau, H. (2014) {Interfacial closure of contacting surfaces.} {\em European Physics Letters} {\bf 107}, 34003.

\bibitem{Springman:2008}
Springman, R. M. and Bassani, J. L. (2008) {Snap transitions in adhesion.} {\em Journal of the Mechanics and Physics of Solids} {\bf 57}, 2358--2380.

\bibitem{Springman:2009}
Springman, R. M. and Bassani, J. L. (2009) {Mechano-chemical coupling in the adhesion of thin-shell structures.} {\em Journal of the Mechanics and Physics of Solids} {\bf 57}, 909--931.

\bibitem{degennes85}
de Gennes, P. G. (1985) {Wetting: statics and dynamics.} {\em Review of Modern Physics} {\bf 104}, 82.
\bibitem{Bonn2009}
Bonn, D, Eggers, J., Indekeu, J., Meunier, J. and Rolley, E. (2009) {Wetting and spreading.} {\em Review of Modern Physics} {\bf 81}, 739.
\bibitem{Snoeijer2013}
Snoejier, J. and  Andreotti, B. (2013) {Moving contact lines: scales, regimes and transitions.}  {\em Annual Review of Fluid Mechanics.} {\bf 45}, 269.
\bibitem{Obreimoff1930}
Obreimoff, J. W. (1930) {The splitting strength of mica}. {\em Proceedings of the Royal Society of London. Series A} {\bf 127}, 290.

\bibitem{Israelachvili2011}
Israelachvili, J. N.  (2011) {\em Intermolecular and surface forces 3rd edition (Elsevier)}. 

 \bibitem{Landau1986}
Landau, L. D. and Lifshitz, E. M. (1986) {\em Theory of elasticity (Elsevier)}.

\bibitem{Batchelor}
Batchelor,  G. K. (1967) {\em Cambridge University Press}, An Introduction to fluid dynamics.

\bibitem{Oron:1997} 
Oron, A., Davis, S. H. and Bankoff, S. G. (1997) {Long-scale evolution of thin liquid films}. {\em Reviews of Modern Physics} {\bf 69}, 931--980.
\bibitem{Gear}
Gear, C. W. (1971) {\em IEEE transactions on circuit theory.} {\bf 18}, 89-95.     

\bibitem{carlson:epl}
Carlson, A., Kim, P., Amberg, G. and Stone, H.A. (2013) {Short and long time drop dynamics on lubricated substrates.} {\em Europhys. Lett.} {\bf 104}, 34008.






\bibitem{timederiv}
{Inserting the scaling factors in (7) into $h_t=-cf'(\eta)H(t)/\ell(t)+cf(\eta)H'(t)-f'(\eta)H(t)\eta \ell'(t)/\ell(t)$ shows that the neglected terms are indeed subdominant as $t \rightarrow \infty $} 

\end{thebibliography}
\end{document}